\documentclass[preprint,aps,prb,showpacs,preprintnumbers,amsmath,amssymb]{revtex4-1}

\usepackage{graphics}
\usepackage{epsfig}
\usepackage{bm}

\begin{document}
\title{Barocaloric and Magnetocaloric Effects in Fe$_{49}$Rh$_{51}$}

\author{Enric Stern-Taulats}

\author{Antoni Planes}
\affiliation{Departament d'Estructura i Constituents de la Mat\`eria, Facultat de F\'isica, Universitat de Barcelona, Mart\'{\i} i Franqu\`es, 1, E-08028 Barcelona. Catalonia.}
\author{Pol Lloveras}
\author{Maria Barrio}
\author{Josep-Llu\'{\i}s Tamarit}
\affiliation{Departament de F\'{\i}sica i Enginyeria Nuclear. ETSEIB. Universitat Polit\`ecnica de Catalunya, Diagonal 647, 08028 Barcelona. Catalonia.}
\author{Sabyasachi Pramanick}
\author{Subham Majumdar}
\affiliation{Department of Solid State Physics, Indian Associaton for the Cultivation of Science. Jadavpur, Kolkata 700 032. India.}
\author{Carlos Frontera}
\affiliation{Institut de Ci\`encia de Materials de Barcelona, Campus UAB, Bellaterra, Catalonia.}

\author{Llu\'{\i}s Ma\~nosa}
\affiliation{Departament d'Estructura i Constituents de la Mat\`eria, Facultat de F\'isica, Universitat de Barcelona, Mart\'{\i} i Franqu\`es, 1. E-08028 Barcelona. Catalonia.}
\date{\today}
\begin{abstract}

We report on calorimetry under applied hydrostatic pressure and magnetic field at the antiferromagnetic (AFM)-ferromagnetic (FM) transition of Fe$_{49}$Rh$_{51}$. Results demonstrate the existence of a giant barocaloric effect in this alloy, a new functional property that adds to the magnetocaloric and elastocaloric effects previously reported for this alloy. All caloric effects originate from the AFM/FM transition which encompasses changes in volume, magnetization and entropy. The strong sensitivity of the transition temperatures to both hydrostatic pressure and magnetic field confers to this alloy outstanding values for the barocaloric and magnetocaloric strengths ($|\Delta S|$/$\Delta p$ $\sim$ 12 J kg$^{-1}$ K $^{-1}$ kbar$^{-1}$ and $|\Delta S|$/$\mu_0\Delta H$ $\sim$ 12 J kg$^{-1}$ K$^{-1}$ T$^{-1}$). Both barocaloric and magnetocaloric effects have been found to be reproducible upon pressure and magnetic field cycling. Such a good reproducibility and the large caloric strengths make Fe-Rh alloys particularly appealing for solid-state cooling technologies at weak external stimuli.

\end{abstract}


\maketitle


\section{Introduction} \label{introduction}

Close to the stoichiometric composition, Fe-Rh solidifies in the CsCl structure ($Pm3m$ space group) and orders 
ferromagnetically below a Curie temperature around 680 K. Upon further cooling, this alloy undergoes a magnetic 
phase transition from a ferromagnetic (FM) to an antiferromagnetic (AFM) state. This transition is first-order, strongly composition dependent and
does not involve breaking in the crystal symmetry. In the FM state
Fe atoms have a $\sim$ 3 $\mu_B$ moment and Rh atoms $\sim$ 1 $\mu_B$, while in the AFM there is no appreciable magnetic moment 
in Rh atoms and Fe atoms have $\sim$ 3 $\mu_B$ moment with opposite sign on successive layers of (111) iron planes  \cite{Shirane1964}. The first-order phase transition involves a significant latent heat (with associated entropy change), and due to a strong magnetostructural coupling the
volume increases by $\sim$ 1\% at the AFM to FM transition.

Although the magnetic transition in Fe-Rh was discovered in the late thirties \cite{Fallot1938}, the physical origin of the mechanisms giving rise to this transition
is still a source of active debate \cite{Mariager2012,Gray2012,Cooke2012,Derlet2012,deVries2013,Staunton2014}. In the recent years there has been a renewed attention in the study of Fe-Rh due to its potential technological interest. On the one hand, the AFM/FM phase transition which occurs at temperatures close to ambient has been found to be useful in thermally assisted magnetic recording devices \cite{Thiele2003}. On the
other hand, the latent heat of the transition gives rise to a large entropy change when the transition is driven by an
external field, which results in giant caloric effects suitable for solid-state refrigeration near room temperature. The present paper is aimed at
investigating Fe-Rh in relation to their caloric properties.

A caloric effect refers to the isothermal entropy change or to the adiabatic temperature change taking place in a material when subjected to an 
external stimulus. Presently, there is intensive research in the study of materials with giant caloric effects near room temperature \cite{Manosa2013,Moya2014}. 
 Materials undergoing ferroic phase transitions are prone to exhibit giant caloric effects \cite{Fahler2012,Moya2014}. 
In these materials changes of ferroic properties are induced by application of the thermodynamically conjugated field
to the ferroic property, giving rise to magnetocaloric \cite{Gschneidner2005,Bruck2005,Planes2009}, and electrocaloric \cite{Mischenko2006,Moya2013} effects for magnetic and electric fields respectively, and to mechanocaloric effects for mechanical stresses. It is worth noting that magnetic and electric properties are described by
 vector quantities but stress (and strain) are second-rank tensor properties. Hence, characterization of mechanocaloric properties involves measurements
 for, at least, two independent stress-tensor components in the case of elastically isotropic materials. Experimentally, mechanocaloric effects are usually studied by measuring the response of a ferroic material to  uniaxial stress and to hydrostatic pressure, and the associated caloric effects are respectively known as elastocaloric and barocaloric. Materials with giant elastocaloric \cite{Bonnot2008,Xiao2013} and barocaloric \cite{Manosa2010,Manosa2011} effects have also recently been reported. 
 
 Caloric effects can be either conventional or inverse depending on whether the  applied field isothermally reduces or increases the material's entropy. While conventional effects are commonly observed, inverse caloric effects have been reported in several ferroic materials \cite{Krenke2005,Sandeman2006,Manosa2011}. These inverse caloric effects are usually related to the existence of an interplay between different ferroic properties of the material \cite{Cakir2012}.

In Fe-Rh a giant magnetocaloric effect was first reported in the early nineties \cite{Nikitin1990,Annaorazov1992}, prior to the seminal work on the giant magnetocaloric effect in Gd-Si-Ge \cite{Pecharsky1997a} that boosted the research in the field. However, Fe-Rh was considered to be of no practical use because the effect was believed to be 
irreversible in an alternating magnetic field and even to disappear after a few cycles \cite{Annaorazov1996,Pecharsky1997b,Annaorazov1996,Franco2012}. Later studies
indicated that reproducibility could be achieved for 5 T fields provided that the sample was subjected to a proper combination of isothermal and adiabatic processes \cite{Manekar2008}. With regards to mechanocaloric effects, studies of the AFM/FM transition under uniaxial stress showed that this alloy also exhibited an elastocaloric effect \cite{Nikitin1992}. Both magnetocaloric and elastocaloric effects are inverse. In Fe-Rh, the symmetry-adapted strain-tensor component describing the structural change accompanying the AFM/FM transition is a dilatational strain (volume change), which couples to hydrostatic pressure. It is therefore expected that the transition will be more sensitive to hydrostatic pressure than to uniaxial stress, and there are indeed evidences of a strong  dependence of the transition temperatures to hydrostatic pressure \cite{Wayne1968,Kushwaha2012}. These facts point to the existence of a large barocaloric effect at the AFM/FM pressure-induced transition. Moreover, since pressure  increases the stability of the low-volume (AFM) phase in such a way that the AFM/FM transition temperature shifts to higher values with increasing pressure, it can be anticipated that the associated barocaloric effect will be conventional. 

In this paper we report on calorimetric measurements under hydrostatic pressure which demonstrate that Fe-Rh alloys do show a giant conventional barocaloric
effect. These experiments have been complemented with calorimetry under applied magnetic field on the same sample, which have enabled us to compare the magnitude 
and reproducibility of the barocaloric effect to those of the inverse magnetocaloric effect in this compound. The paper is organised as follows: Section \ref{experimental} is devoted to the experimental details, and results are presented in Section \ref{results}. In Section \ref{discussion} we briefly discuss the obtained data and the main conclusions of the work are compiled in Section \ref{summary}.


\section{Experimental Details} \label{experimental}

A polycrystalline sample of nominal composition Fe$_{49}$Rh$_{51}$ was prepared by arc melting the pure metals under argon atmosphere in a water-cooled Cu crucible. For homogeneity, the sample was remelted several times turning the ingot back to back. Next, the ingot was vacuum sealed in a quartz tube and annealed at 1100 $^o$C for 72 h followed by a furnace cooling to room temperature. From the ingot a 3.3 mm $\times$ 3.0 mm $\times$ 5.6 mm parallelepiped sample (504.36 mg mass) was cut for calorimetric measurements under pressure. A 1 mm diameter and $\sim$ 2 mm length hole was drilled to that sample to host the thermocouple. A second thinner sample (190.1 mg mass) with 1.1 mm thickness and a flat surface of 5.5 mm $\times$ 6.4 mm was cut for calorimetric measurements under applied magnetic field.

Calorimetric measurements under hydrostatic pressure were carried out by means of a custom-built calorimeter described in \cite{Manosa2010}. The thermal signal was measured by a chromel-alumel thermocouple embedded into the sample. Calorimetric runs are performed by scanning temperature at typical rates 2 K min$^{-1}$ (heating) and 1 K min$^{-1}$ (cooling) while hydrostatic pressure is kept constant. From the calorimetric curves at selected values of pressure, the entropy change (referenced to a given state at $T_0$ above the phase transition) is computed as described in \cite{Manosa2011}. 

Calorimetric measurements under magnetic field were carried out by means of a custom-built high-sensitivity differential scanning calorimeter (DSC) described in  \cite{Emre2013}. That device allows both  isofield measurements performed by scanning the temperature (typical rates $\pm  0.5 $ K min$^{-1}$) and also isothermal measurements performed by scanning the magnetic field (typical rates $\pm 0.16 $ T min$^{-1}$). From these measurements quasi-direct (isofield data) and direct (isothermal data) computations of the entropy change are performed as described in \cite{Emre2013}. 

Complementary magnetization measurements were carried out in a physical property measurement system (PPMS, Quantum Design).

\section{Results} \label{results}

Figure 1 shows calorimetric curves (sweeping temperature) at selected values of hydrostatic pressure without magnetic field (left panels) and
at selected values of magnetic field at atmospheric pressure (right panels). The magnetostructural transition gives rise to a large  exothermal peak on cooling (lower panels)
and endothermal peak on heating (upper panels). The transition is sharp (it spreads over less than 5 K) and takes place with a thermal hysteresis width
of $\sim$ 10 K. The transition shifts to higher temperatures with increasing pressure while it shifts to lower temperatures with increasing
magnetic field. This behaviour is consistent with pressure stabilizing the lower volume AFM state and magnetic field stabilizing
the larger magnetization FM phase. 

The temperature dependence of magnetization ($M(T)$) measured during cooling and heating across the AFM/FM transition is shown in Fig. 2 for selected values of the magnetic field. It is found that $M(T)$ remains almost temperature independent in both AFM and FM phases and sharply changes at the AFM to FM transition on heating and at the FM to AFM
transition on cooling, with a thermal hysteresis which compares well to that derived from calorimetric data.

In Fig. 3({\it a}) we show the temperatures of the calorimetric peaks  for ($T^c$) forward (FM to AFM) and ($T^h$) reverse (AFM to FM) transitions plotted as
a function of applied pressure and magnetic field. Data exhibit a very good linear behaviour with slopes $dT^c/dp$= 6.4 K kbar$^{-1}$; and $dT^c/\mu_0dH$=  -9.6 K T$^{-1}$
for the FM to AFM transition and $dT^h/dp=$ 5.4 K kbar$^{-1}$ and $dT^h/\mu_0dH$= -9.7 K T$^{-1}$ for the AFM to FM transition respectively. Thermal hysteresis is not significantly affected by the magnetic field and it marginally decreases with increasing pressure. At much higher pressures ($>$ 50 kbar) the pressure dependence of the transition temperature is expected to weaken as the sample approaches the triple point \cite{Wayne1968}.

By numerical integration of calorimetric curves as described in \cite{Manosa2011} and \cite{Emre2013}  we have obtained
the entropy ($\Delta S_t$) and enthalpy ($\Delta h_t$)  changes corresponding to the AFM/FM transition. Averaged (heating and cooling) values are shown in Figs. 3({\it b}) and 3({\it c}) respectively. The values at zero field and atmospheric pressure 
$\Delta S_t$ = 12.5 $\pm$ 1 J kg$^{-1}$ K$^{-1}$ and $\Delta h_t$ = 3900 $\pm$ 150 J kg$^{-1}$ are in agreement with previously reported data \cite{Richardson1973}. It is worth noting that for the studied magnetostructural transition, $\Delta h_t$ is to a very good approximation the energy difference ($\Delta E$) between AFM and FM phases.

\begin{figure}[h]
\includegraphics[width=8cm]{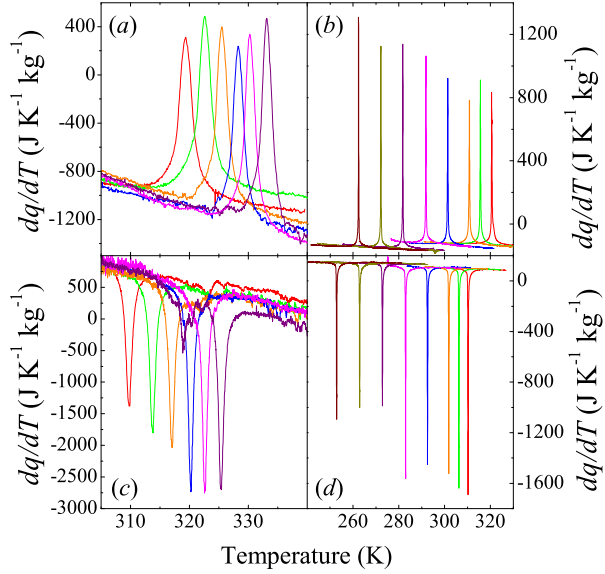}
\caption{Calorimetric curves recorded at selected values of hydrostatic pressure [({\it a}) and ({\it c})] and magnetic field [({\it b}) and ({\it d})]. Positive values
correspond to endothermal processes obtained during heating [({\it a}) and ({\it b})] while negative values correspond to exothermal processes recorded
during cooling [({\it c}) and ({\it d})]. In the left panels, data correspond to pressures (from left to right) of $p$=0, 0.6, 1.1, 1.6, 2.0, and 2.5 kbar. In the right panels, data correspond to magnetic fields (from right to left) of $\mu_0 H$= 0, 0.5, 1, 2, 3, 4, 5 and 6T.} \label{fig1}
\end{figure}

\begin{figure}[h]
\includegraphics[width=8cm]{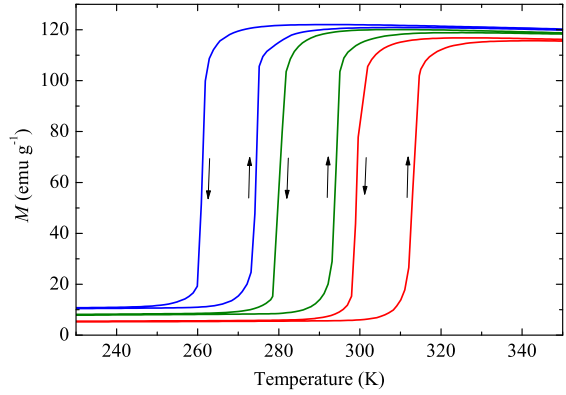}
\caption{Temperature dependence of the magnetization for cooling and heating runs under applied magnetic field. From right to left data correspond to 1, 3 and 5 T.}\label{fig2}
\end{figure}

\begin{figure}[h]
\includegraphics[width=8cm]{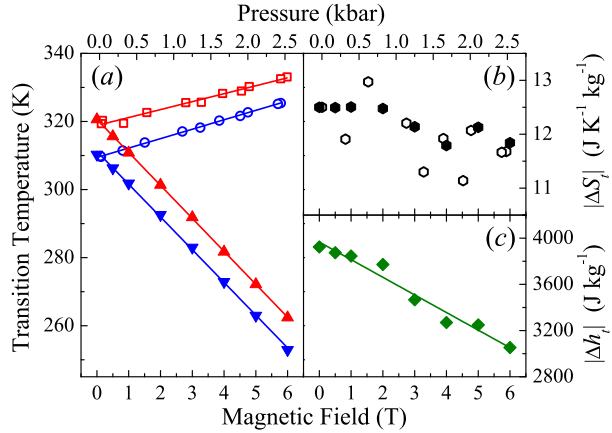}
\caption{({\it a}) Temperature of the calorimetric peak as a function of pressure (open symbols) and magnetic field (solid symbols). Blue symbols (down triangles and
circles) stand for cooling runs while red symbols (up triangles and squares) stand for heating runs. Solid lines are linear fits to the data. ({\it b}) Averaged values (between heating and cooling) for the transition entropy change as a function of pressure (open symbols) and magnetic field (solid symbols).({\it c}) Averaged values (between heating and cooling) for the transition enthalpy change as a function of magnetic field. The line is a linear fit to the data.} \label{fig3}
\end{figure}

Calorimetric curves at selected values of hydrostatic pressure and magnetic field enable us to determine the isothermal entropy changes (quasi-direct method) 
associated with the barocaloric and magnetocaloric effects. Results are shown in Fig. 4. Barocaloric effect has been found to be conventional and magnetocaloric effect is inverse. That is, while isothermal application of pressure reduces the total entropy, magnetic field increases the total entropy of the alloy. The conventional
and inverse nature of barocaloric and magnetocaloric effects are consistent with pressure stabilizing the low temperature AFM phase and magnetic field stabilizing the
high temperature FM phase.  

The pressure-induced entropy change and magnetic field-induced entropy change increase in magnitude as pressure and magnetic field increase respectively, until
a saturation value is reached. This behaviour gives rise to a plateau in the $\Delta S$ vs $T$ curves. Both barocaloric and magnetocaloric effects saturate to the 
same value which is coincident with the transition entropy change $|\Delta S_t|$.  This result shows that both caloric effects have the same origin, and
the giant values for the pressure-induced and field-induced entropy changes are a consequence of the magnetostructural transition which involves a 
large entropy change ($\Delta S_t$). The saturation value for the barocaloric and magnetocaloric effects is reached for low values of pressure and magnetic field, as
illustrated in Fig. 5({\it a}) which shows $|\Delta S|_{max}$, the absolute value of the maximum in the $\Delta S$ vs $T$ curves depicted in Fig. 4, as a function of pressure and magnetic field. 

\begin{figure}[h]
\includegraphics[width=8cm]{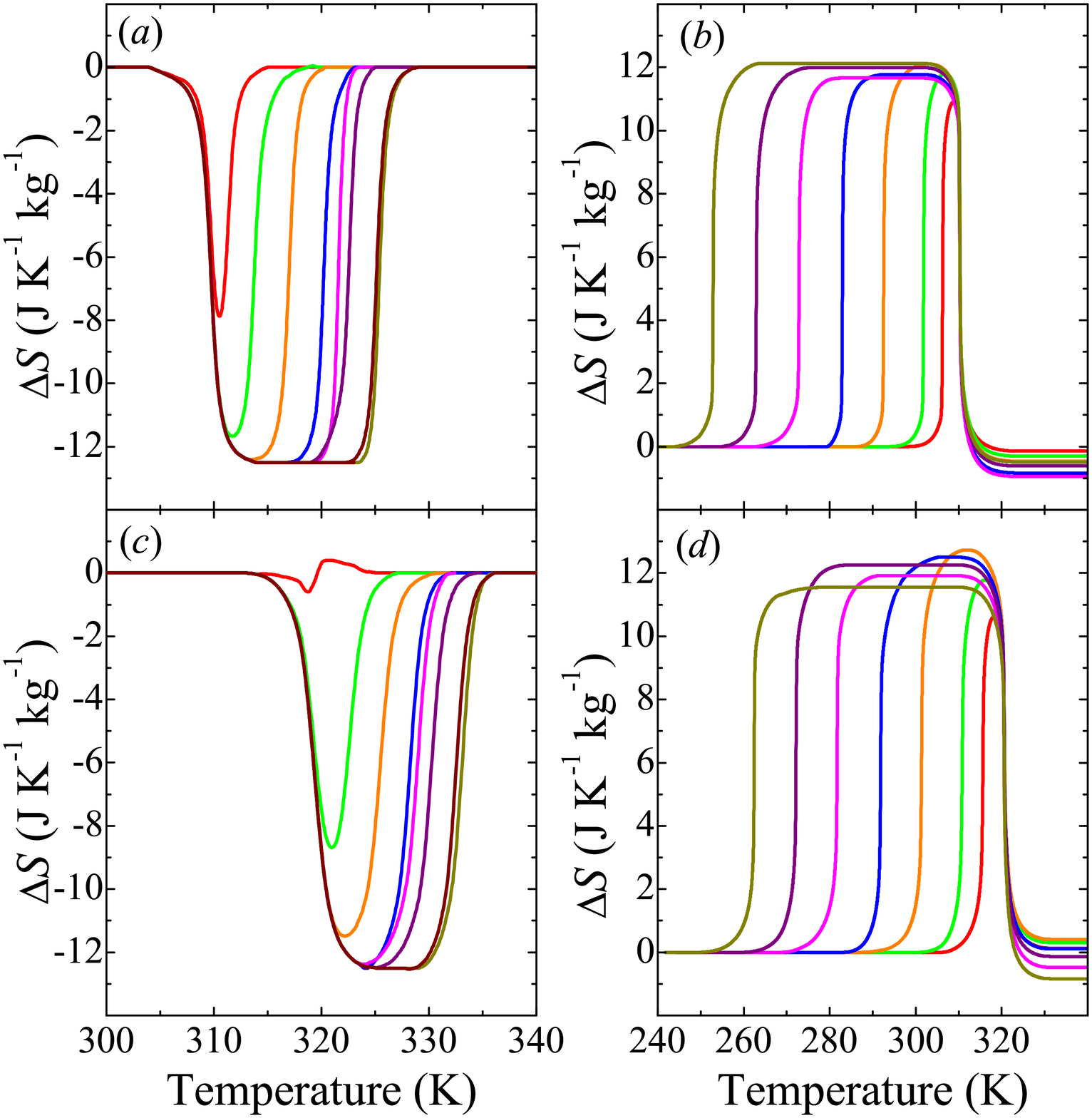}
\caption{({\it a}) and ({\it c}) Pressure-induced entropy change (barocaloric effect) and ({\it b}) and ({\it d}) magnetic field-induced entropy change (magnetocaloric
effect) as a function of temperature for selected values of hydrostatic pressure and magnetic field. Upper panels ({\it a}) and ({\it b}) correspond to cooling runs and lower panels ({\it c}) and ({\it d}), to heating runs. Data for barocaloric effect correspond to pressures (from left to right) of $p$ = 0.3, 0.6, 1.1, 1.6, 1.9, 2.0, 2.4, and 2.5 kbar. Data for magnetocaloric effect correspond to magnetic fields (from right to left) of $\mu_0 H$=  0.5, 1, 2, 3, 4, 5, and 6 T.} \label{fig4}
\end{figure}
The performances of a given material for solid-state refrigeration are typically analysed in terms of the relative cooling power ($RCP$) which provides
an estimate about the amount of heat that can be transferred in a field cycle between cold and hot reservoirs, and is defined as $RCP = |\Delta S|_{max} \times \delta T_{FWHM}$, where  $\delta T_{FWHM}$ is the temperature width at half maximum of the $\Delta S$ vs $T$ curves (Fig. 4). These values are shown in Fig. 5({\it b}) for the barocaloric and
 magnetocaloric effects.

\begin{figure}[h]
\includegraphics[width=8cm]{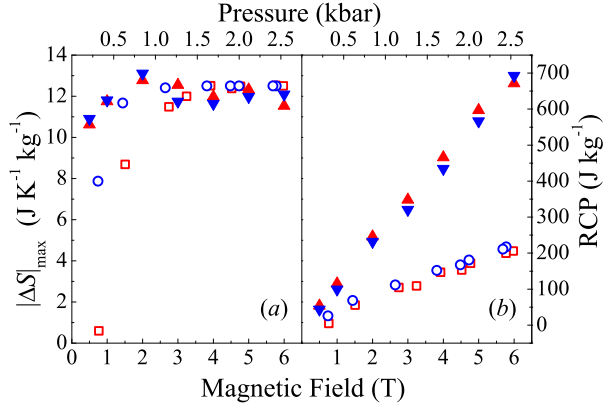}
\caption{Absolute value of the maximum entropy change ({\it a}) and relative cooling power ({\it b}) as a function of pressure (open symbols) and magnetic field (solid symbols). Blue symbols (down triangles and circles) stand for cooling runs and red symbols (up triangles and squares) stand for heating runs. } \label{fig5}
\end{figure}

\section{Discussion} \label{discussion}

We have shown that in addition to the already reported magnetocaloric and elastocaloric effects, Fe-Rh alloys also exhibit a barocaloric effect associated with the volume change at the AFM/FM phase transition which enables the transition to be driven by hydrostatic pressure. The maximum pressure-induced entropy change value found for Fe$_{49}$Rh$_{51}$ 
($|\Delta S|$ = 12.5 J kg$^{-1}$ K$^{-1}$) compares well with the values reported for other giant barocaloric materials \cite{Manosa2010,Manosa2011,Yuce2012}. Interestingly, such a maximum isothermal entropy change is achieved for relatively low pressures. This establishes Fe-Rh to be a material with a large barocaloric strength ($|\Delta S|/\Delta p$) of  $\sim 12$ J K$^{-1}$ kg$^{-1}$ kbar$^{-1}$. Indeed, the magnetocaloric strength ($|\Delta S|/\mu_0 \Delta H$)  $\sim$ 12 J K$^{-1}$ kg$^{-1}$ T$^{-1}$ is also one of the largest reported so far among giant magnetocaloric materials  \cite{Franco2012}. These outstanding values for the caloric strengths arise from the sharpness of the transition and the strong sensitivity of the transition temperatures to both pressure and magnetic field. 

It is worth comparing the caloric response of Fe-Rh to hydrostatic pressure and uniaxial stress. The sensitivity of the transition temperature to hydrostatic pressure ($dT/dp$ $\simeq$ 6$\times$ 10$^{-8}$ K Pa$^{-1}$) is about three times larger (in absolute value) than to uniaxial stress ($dT/d\sigma$ $\simeq$ - 2 $\times$ 10$^{-8}$ K Pa$^{-1}$) \cite{Nikitin1992} which indicates a larger barocaloric effect than the previously reported elastocaloric one. Although there are no entropy values available for the elastocaloric effect, the estimated uniaxial stress value to induce the full AFM/FM transition (resulting in the saturation value for $\Delta S$) is $\sim$ 300 MPa which renders a lower elastocaloric strength $|\Delta S|$/$\Delta \sigma$ $\sim$ 4 $\times$ 10$^{-8}$ J kg$^{-1}$ K$^{-1}$ Pa$^{-1}$ (= 4 J kg$^{-1}$ K$^{-1}$ kbar$^{-1}$).

\begin{figure}[h]
\includegraphics[width=6.5cm]{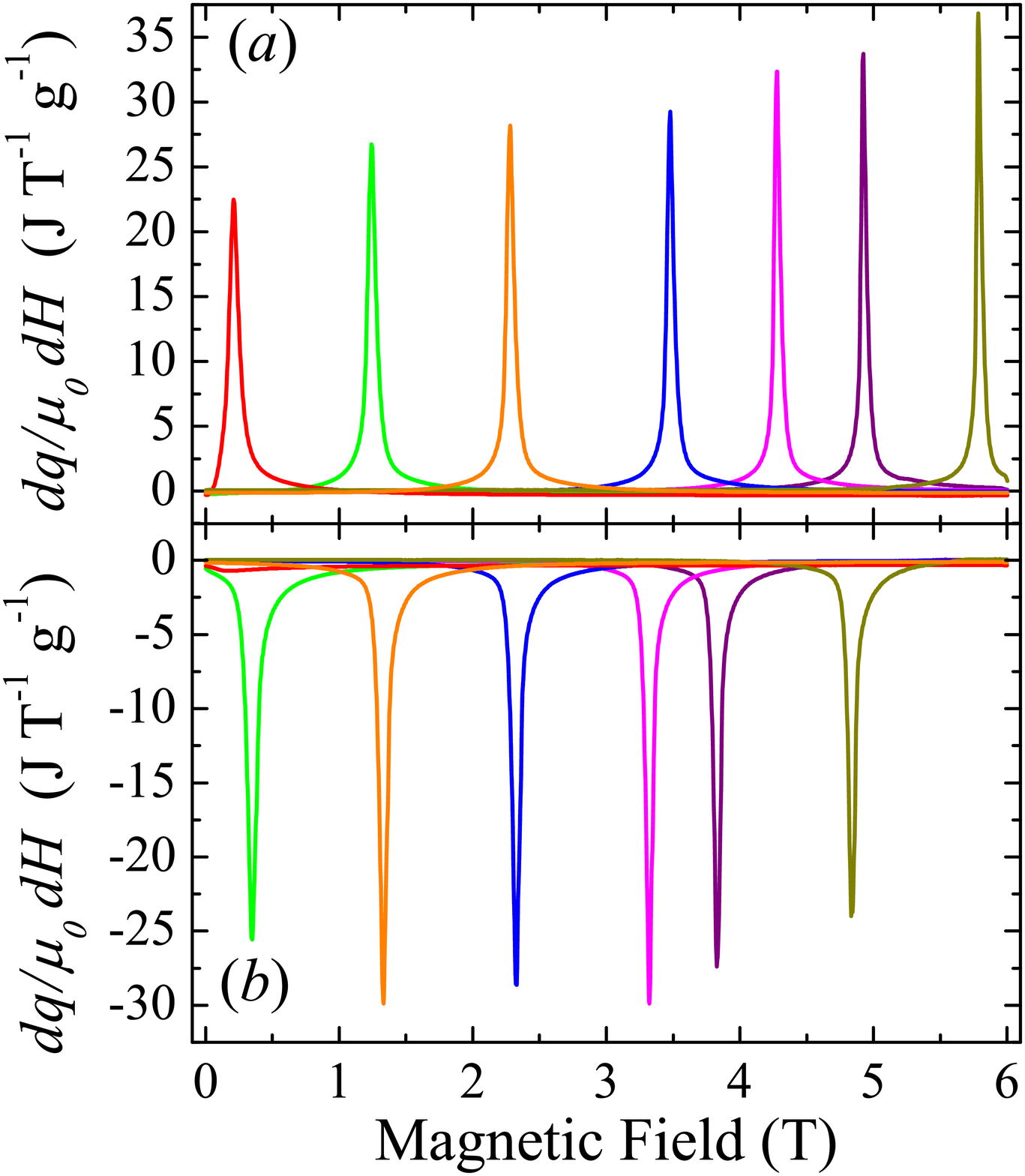}
\caption{Thermograms obtained upon field scanning  at selected values of temperature. From right to left $T$ = 264.7, 272.6, 279.3, 289.1, 298.3, 307.4 and 315.5 K. Positive signals (endothermal peaks) are recorded upon application of magnetic field from 0 to 6T and negative signals (exothermal peaks) are recorded upon removal of the field from 6T to 0.} \label{fig6}
\end{figure}

\begin{figure}[h]
\includegraphics[width=8cm]{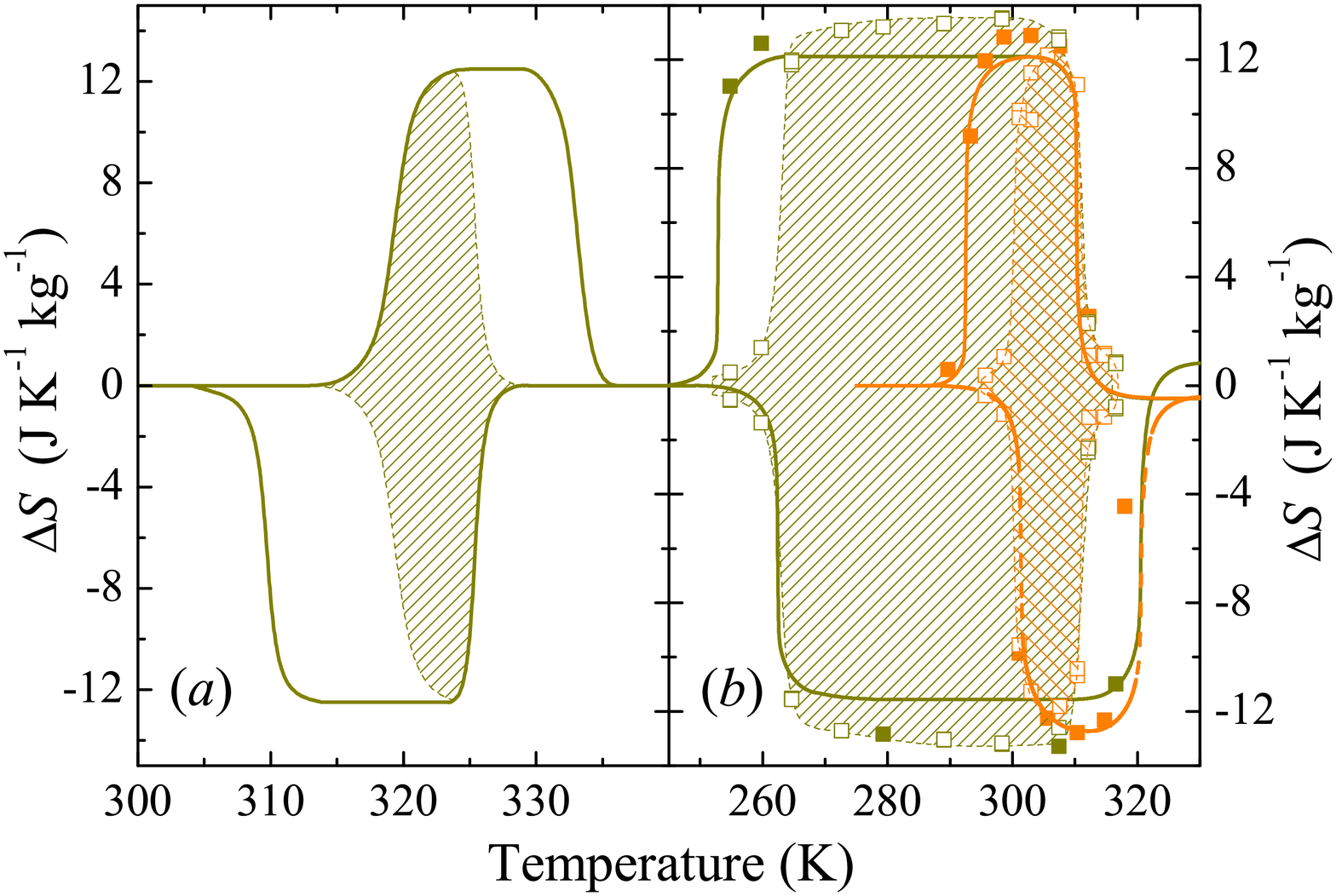}
\caption{Barocaloric ({\it a}) and magnetocaloric ({\it b}) effects. For convenience, data on heating have been plotted as positive in the barocaloric effect and
as negative in the magnetocaloric effect. In ({\it a}) lines correspond to the barocaloric effect at 2.5 kbar. In ({\it b}) symbols represent the field-induced entropy values computed from isothermal thermal curves (direct method). Solid symbols correspond to the first application (or removal) of the field and open symbols correspond to successive field cycling. Dashed lines are a guide to the eye and solid lines are the values computed from isofield thermal curves (quasi-direct method). Orange symbols and lines correspond to 2 T magnetic field and green symbols and lines, to 6 T. In all cases the shaded areas indicate the region of reversibility.} \label{fig7}
\end{figure}

\begin{figure}[h]
\includegraphics[width=7cm]{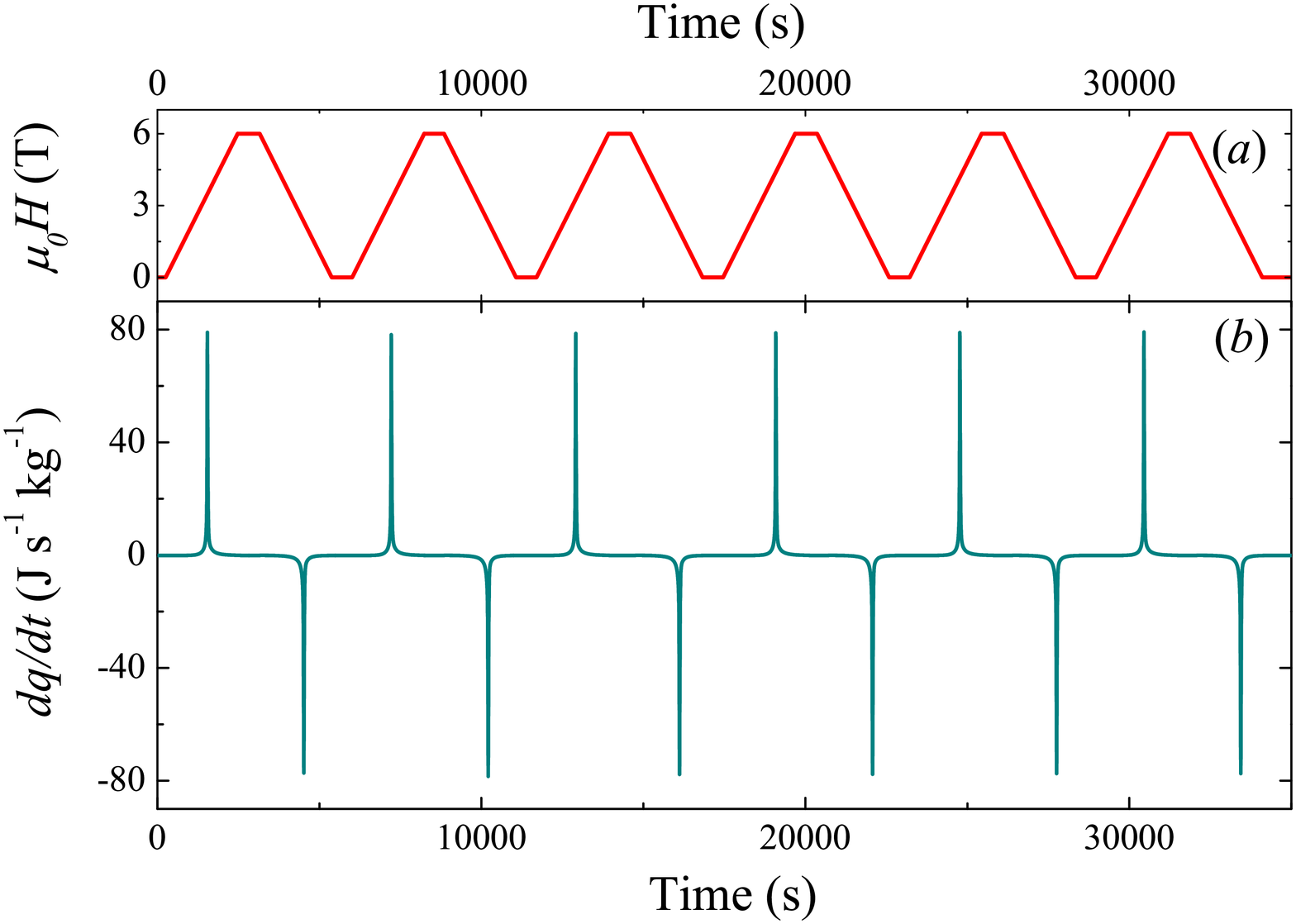}
\caption{(a) Applied magnetic field and (b) recorded isothermal calorimetric curves as a function of time. Data correspond to isothermal field cycles between 0 and 6 T at a temperature of 289 K.} \label{fig8}
\end{figure}

As previously mentioned, the reproducibility in the magnetocaloric effect has been a controversial issue \cite{Annaorazov1996,Pecharsky1997b,Franco2012,Manekar2008}. Isothermal DSC with magnetic field   enables direct determination of the magnetic field-induced entropy change and it is a unique tool to study the reproducibility of the magnetocaloric effect upon field cycling \cite{Casanova2005,SternTaulats2014}. We have performed calorimetric measurements at selected values of the temperature while magnetic field was swept. The measurement protocol is described in detail in \cite{Emre2013}. Fig. 6 shows the calorimetric signal as a function of magnetic field, recorded on the first application (upper curves) and first removal (lower curves) of a 6T field at selected values of temperature. Increasing the field drives the sample from AFM to FM phases with the absorption of latent heat (endothermal process) while the sample transforms from FM to AFM upon removal of the field and releases the latent heat (exothermal process). By taking the peak value as the transition field at each temperature we obtain a field dependence of the transition temperature that perfectly matches data from isofield measurements shown in Fig. 3.

Integration of isothermal calorimetric curves provide a direct determination of  the field-induced entropy change ($\Delta S$). 
Results for 2 T and 6 T are plotted in Fig. 7({\it b}) as solid symbols, and are compared to the quasi-direct determination from isofield calorimetric data (solid lines) described in the preceding section. There is good agreement between the two sets of data. The reproducibility has been studied by isothermal calorimetric measurements under
cyclic variation of magnetic field. An illustrative example of the recorded calorimetric signals  upon isothermal succesive
magnetic field cycles between 0 and 6 T  is shown in Fig. 8.  The good reproducibility exhibited by calorimetric curves demonstrates an excellent reversibility of the magnetocaloric effect. $\Delta S$ values are computed from numerical integration of these curves and are constant upon successive field cycling within experimental error. Data for all studied temperatures at 2 and 6 T are indicated as open symbols in Fig. 7({\it b}).

Reversibility of a given caloric effect is expected to be restricted within a certain temperature range which depends on the magnitude of the applied external field. This region can be determined from experiments carried out both on cooling and heating \cite{Emre2013} and is indicated as a shaded area in Fig. 7.  In the case of a conventional caloric 
effect (as the barocaloric effect here) this region is bounded by the transition temperature of the reverse (AFM to FM) transition at atmospheric pressure and the transition temperature of the forward (FM to AFM) transition under applied pressure. For an applied pressure of 2.5 kbar the region where barocaloric effect will be reproducible extends from 319 to 325 K. For an inverse caloric effect (as the magnetocaloric here) the reversibility region is bounded by the reverse (AFM to FM) transition temperature under applied field and the forward (FM to AFM) transition temperature at zero field. The magnetocaloric effect is reversible within 294 and 306 K for 2 T and within 257 and 306 K for 6 T.

The isothermal entropy changes associated with the barocaloric and magnetocaloric effects saturate for pressures $\sim$ 1 kbar and fields $\sim$ 1 T to a value which coincides with the total transition entropy change (see Fig. 5). Such a tendency towards saturation has not been found in other giant magnetocaloric materials. For instance, in Gd-Si-Ge, the entropy change shows a monotonous increase with increasing magnetic field \cite{Casanova2004} while in some magnetic shape memory alloys the entropy change increases up to a maximum value and decreases upon further increasing magnetic field \cite{Recarte2010}. These different behaviours can be understood by taking into consideration that the entropy change contains contributions from the latent  heat (transition entropy change) and also intrinsic contributions from both high temperature and low temperature phases \cite{Casanova2002}. For the particular case of Fe-Rh the fact that magnetocaloric and barocaloric data saturate to a value which coincides with the transition entropy change indicates that the intrinsic magnetic and elastic contributions of the AFM and FM phases are small. 
In the magnetic case, the importance of these intrinsic contributions is given by the value $(\partial M/\partial T)_H$ in each phase. As shown in Fig. 2, $M$ vs $T$ curves are almost flat in both AFM and FM phases, for different values of magnetic field, and the estimated intrinsic contributions to $\Delta S$ are one order of magnitude lower than that from the magnetostructural transition. In the FM state, such a weak temperature dependence is due to an almost saturated FM order since the transition takes place well below the Curie point. In the AFM, neither temperature nor magnetic field significantly affect the AFM order. This could be attributed to a large magnetocrystalline anisotropy.

For the barocaloric effect, the intrinsic contribution is given by 

\begin{equation}
\Delta  S = \int_0^p  \beta v dp 
\end{equation}

\noindent Where $\beta$ is the thermal expansion and $v$, the specific volume. By using reported data for $\beta$ and $v$ \cite{Cooke2012}, we estimate that for $p$=2.5 kbar this contribution amounts to $\sim$ 0.6 Jkg$^{-1}$K$^{-1}$ in the AFM phase and to $\sim$ 0.5 Jkg$^{-1}$K$^{-1}$ in the FM phase. These values are small compared to the contribution arising from the transition entropy change. 

Although a dependence of the transition entropy change to pressure and magnetic field falls within experimental errors, data show a tendency to slightly decrease with increasing pressure and magnetic field [Fig. 3({\it b})]. Previous indirect measurements from magnetization data indicated a larger decrease in the isothermal $\Delta S$ with increasing magnetic field \cite{Annaorazov1996}. By contrast, the energy difference between AFM and FM phases shows a marked decrease as magnetic field increases [Fig. 3({\it c})], with an average rate d$\Delta E/dH \simeq$ - 150 Jkg$^{-1}$T$^{-1}$. It is worth noting that recent adiabatic calorimetry experiments \cite{Cooke2012} did not find any magnetic field dependence of the specific heat of AFM and FM samples which would point to a magnetic field independent transition enthalpy change. Probably the different behaviour must be ascribed to a highest sensitivity of differential scanning calorimetry in determining enthalpy changes, and also to the fact that specific heat data correspond to two different samples while present experiments are carried out on a single specimen. It is acknowledged that the features of the AFM/FM transition in Fe-Rh are extremely sensitive to composition \cite{Staunton2014}.


Fe-Rh exhibits a conventional barocaloric effect and an inverse magnetocaloric effect. Similar behaviour is also present in
several Ni-Mn-based magnetic shape memory alloys \cite{Manosa2010,Manosa2014}. In both cases the giant caloric effects are associated with a first-order phase transition which involves a decrease in magnetization and volume when the sample transforms from the high-temperature to the low-temperature phases. The total entropy of the alloy also decreases at this phase transition.
There are, however, significant differences in the physical origins of the caloric effects 
between the two families of alloys when considering the different contributions to the entropy change which drives the magnetostructural phase transition. In Fe-Rh there is still some controversy on whether the major contribution to the entropy change arises from conduction electrons or from magnetic moments. While x-ray photoemission \cite{Gray2012} and Hall-effect \cite{deVries2013} measurements indicate significant changes in the electron density of states at the AFM/FM 
transition, recent specific heat measurements \cite{Cooke2012} suggest that the electronic contribution is small and the entropy difference at the transition is dominated by magnetic fluctuations. In any case, both electronic and magnetic contributions are lower in the AFM than in the FM phase, thus stabilizing the low-temperature phase. By contrast, the entropy associated with the lattice is larger in the AFM phase than in the FM one, due to the fact that AFM phase is elastically softer than the FM one, as results from the lower values for both longitudinal and transverse Debye temperatures of the AFM phase. Therefore in Fe-Rh the magnetostructural transition
is driven by an excess in electronic and  magnetic entropy while the lattice entropy opposes to the transition.  By contrast, in magnetic shape memory alloys the low-temperature phase is a short range antiferromagnetic phase \cite{Aksoy2009}, with a magnetic entropy larger than that of the high temperature FM phase, but the entropy arising from lattice 
vibrations is lower in the close-packed phase than in the FM cubic phase \cite{Recarte2012}. Such an excess of vibrational entropy arises from a low-energy TA$_2$ transverse phonon branch in the open cubic phase and is responsible for driving the magnetostructural (martensitic) transition \cite{Planes2001}. The electron contribution is only playing a minor role in martensitic transitions \cite{Siewert2012}. Therefore, while the transition in Fe-Rh is magnetically driven, with magnetization being the primary ferroic property, in magnetic shape memory alloys the martensitic transition is vibrationally driven, with a shear strain as a primary ferroic property.

\section{Summary and conclusions} \label{summary}

By means of calorimetry under hydrostatic pressure we have shown that Fe$_{49}$Rh$_{51}$ exhibits
a giant barocaloric effect. This new functional property adds to the already reported magnetocaloric and elastocaloric 
effects in this alloy. All these caloric effects share the same physical origin which is the occurence of a first order AFM/FM phase transition which
encompasses a significant entropy change. Actually, this  transition entropy change ($|\Delta S_t| = 12.5 \pm 1$ J kg$^{-1}$ K$^{-1}$) represents
the upper bound for the pressure-induced and magnetic field-induced entropy changes.

The reproducibility of the magnetocaloric effect has been studied by a direct determination of the field-induced entropy
change from isothermal calorimetric measurements. The comparison between direct and quasi-direct methods at the
magnetocaloric effect has enabled us to assess also the reproducibility of the barocaloric effect from the quasi-direct data. We have found that for a field of 2 T the magnetocaloric effect is perfectly reproducible upon field cycling. This reproducibility is restricted within the temperature range 294-306 K and is increased to 257-306 K for a field of 6 T. The barocaloric effect is estimated to be reversible upon pressure cycling in the temperature range 319-325 K for applied pressures of 2.5 kbar.

Materials with cross-response to more than one external field are particularly interesting \cite{Moya2014}. In Fe-Rh application of hydrostatic pressure enhances the stability of the AFM phase and shifts the $H-T$ transition line to higher temperature values, and application of magnetic field enhances the stability of the FM phase and shifts the $p-T$ transition line to lower temperature values. Interestingly, such an opposite sensitivity of the transition to pressure and magnetic field has been proved to be useful in reducing the hysteresis of magnetostructural phase transitions by a proper combination of pressure and magnetic field \cite{Liu2012}.

The energy difference between AFM and FM phases has been found to decrease with increasing magnetic field. Present results provide reproducible experimental data which we expect will encourage the development of theoretical models that include the effect of magnetic field in the computation of both  energy and entropy values for the different phases involved in the transition. The combination of reliable experimental data and theoretical modelling should help in the understanding of the role played by the different contributions (electronic, magnetic and structural) in driving the AFM/FM transition in Fe-Rh alloys.  

The sharpness of the transition together with the strong sensitivity of the transition to the external fields results in 
barocaloric and magnetocaloric strengths which compare favourably to those reported for other giant magnetocaloric and barocaloric
materials. As a consequence of such  large strengths, Fe$_{49}$Rh$_{51}$ achieves its maximum isothermal entropy change at very
low values of hydrostatic pressure and magnetic field. This fact, added to the aforementioned good reproducibility makes
this alloy particularly interesting in cooling applications where the external stimuli need to be restricted to low values.

\begin{acknowledgments} We acknowledge financial support from CICyT (Spain),  projects No. MAT2013-40590-P and FIS2011-24439, and Joint Indo-Spanish project, 
DGICyT (Spain), project No. PRI-PIBIN-2011-0780 and DST (India) project No. DST/INT/P-39/11. E. Stern-Taulats acknowledges support from AGAUR (Catalonia). 
\end{acknowledgments}


\begin{thebibliography}{99}

\bibitem{Shirane1964} G. Shirane, R. Nathans, C.W. Chen, Phys. Rev. {\bf 134}, A1547 (1964).

\bibitem{Fallot1938} M. Fallot, Ann. Phys. (Paris) {\bf 10}, 291 (1938).

\bibitem{Mariager2012} S. O. Mariager {\it et al.}, Phys. Rev. Lett. {\bf 108}, 087201 (2012).

\bibitem{Gray2012} A. X. Gray {\it et al.}, Phys. Rev. Lett. {\bf 108}, 257208 (2012).

\bibitem{Cooke2012} D. W. Cooke, F. Hellman, C. Baldasseroni, C. Bordel, S. Moyerman, and E.E. Fullerton,
Phys. Rev. Lett. {\bf 109}, 255901 (2012).

\bibitem{Derlet2012} P. M. Derlet, Phys. Rev. B {\bf 85}, 174431 (2012).

\bibitem{deVries2013} M. A. de Vries, M. Loving, A. P. Mihai, L. H. Lewis, D. Heiman, and C.H. Marrows,
New. J. Phys. {\bf 15}, 013008 (2013).

\bibitem{Staunton2014} J. B. Staunton, R. Banerjee, M. dos Santos Dias, A. Deak, and L. Szunyogh, Phys. Rev. B {\bf 89}, 054427 (2014).

\bibitem{Thiele2003} J. U. Thiele, S. Maat, and E.E. Fullerton, Appl. Phys. Lett. {\bf 82}, 2859 (2003).

\bibitem{Manosa2013} L. Ma\~nosa, A. Planes, and M. Acet, J. Mater. Chem. A, {\bf 1}, 4925 (2013).

\bibitem{Moya2014} X. Moya, S. Kar-Narayan, and N. D. Mathur, Nature Mater. {\bf 13}, 439 (2014).

\bibitem{Fahler2012} S. F\"ahler {\it et al}. Adv. Engn. Mat. {\bf 14}, 10 (2012).

\bibitem{Gschneidner2005}  K. A. Gschneidner, V. K. Pecharsky, and A. O. Tsokol, Rep. Prog. Phys. {\bf 68}, 1479 (2005).

\bibitem{Bruck2005} E. Br\"uck, J. Phys. D: Appl. Phys. {\bf 38}, R381 (2005).

\bibitem{Planes2009} A. Planes, L. Ma\~nosa, and M. Acet, J. Phys.: Condens. Matter {\bf 21}, 233201 (2009).

\bibitem{Mischenko2006} A. S. Mischenko, Q. Zhang, J. F. Scott, R. W. Wathmore, and N. D. Mathur, Science {\bf 311}, 1270 (2006).

\bibitem{Moya2013} X. Moya, E. Stern-Taulats, S. Crossley, D. Gonz\'alez-Alonso, S. Kar-Narayan, A. Planes, L. Ma\~nosa, and N. D. Mathur, Adv. Mater. {\bf 25}, 1360 (2013). 

\bibitem{Bonnot2008} E. Bonnot, R. Romero, L. Ma\~nosa, E. Vives, and A. Planes, Phys. Rev. Lett. {\bf 100}, 125901 (2008).

\bibitem{Xiao2013} F. Xiao, T. Fukuda, and T. Kakeshita, Appl. Phys. Lett. {\bf 102}, 161914 (2013).

\bibitem{Manosa2010} L. Ma\~nosa, D. Gonz\'alez-Alonso, A. Planes, E. Bonnot, M. Barrio, J. L. Tamarit, S. Aksoy, and M. Acet, Nature Mater. {\bf 9}, 478 (2010).

\bibitem{Manosa2011} L. Ma\~nosa, D. Gonz\'alez-Alonso, A. Planes, M. Barrio, J. L. Tamarit, I. S. Titov, M. Acet, A. Bhattacharyya, and S. Majumdar, Nature Comm. {\bf 2}, 595 (2011).

\bibitem{Krenke2005} T. Krenke, E. Duman, M. Acet, E. F. Wassermann, X. Moya, L. Ma\~nosa, and A. Planes, Nature Mater. {\bf 4}, 450 (2005).

\bibitem{Sandeman2006} K.G. Sandeman, R. Daou, S. \"Ozcan, J. H. Durrell, N. D. Mathur, and D. J. Fray, Phys. Rev. B {\bf 74}, 224436 (2006).

\bibitem{Cakir2012} O. Cakir, and M. Acet, Appl. Phys. Lett. {\bf 100}, 202404 (2012).

\bibitem{Nikitin1990} S. Nikitin, G. Myalikgulyev, A. M. Tishin, M. P. Annaorazov, K. A. Asatryan, and A. L. Tyurin,
Phys. Lett. {\bf 148}, 363 (1990).

\bibitem{Annaorazov1992} M. P. Annaorazov, K. A. Asatryan, G. Myalikgulyev, S. A. Nikitin, A. M. Tishin, and A. L. Tyurin.
Cryogenics {\bf 32}, 867 (1992).

\bibitem{Pecharsky1997a} V. K. Pecharsky, and K. A. Gschneidner, Phys. Rev. Lett. {\bf 78}, 4494 (1997).

\bibitem{Franco2012} V. Franco, J. S. Bl\'azquez, B. Ingale, and A. Conde, Annu. Rev. Mater. Res. {\bf 42}, 305 (2012).

\bibitem{Annaorazov1996} M. O. Annaorazov, S. A. Nikitin, A. L. Tyurin, K. A. Asatryan,  and A. K. Dovletov, J. Appl. Phys. {\bf 79}, 1689 (1996).

\bibitem{Pecharsky1997b} V. K. Pecharsky, and K. A. Gschneidner, Appl. Phys. Lett. {\bf 70}, 3299 (1997).

\bibitem{Manekar2008} M. Manekar, and S.B. Roy, J. Phys. D: Appl. Phys. {\bf 41}, 192004 (2008).

\bibitem{Nikitin1992} S. Nikitin, G. Myalikgulyev, M. P. Annaorazov, A. L. Tyurin, R. W. Myndev, and S. A. Akopyan,
Phys. Lett. {\bf 171}, 234 (1992).

\bibitem{Wayne1968} R. C. Wayne, Phys. Rev. {\bf 170}, 523 (1968).

\bibitem{Kushwaha2012} P. Kushwaha, P. Bar, R. Rawat, and P. Chaddah, J. Phys.: Condens. Matter. {\bf 24}, 096005 (2012).

\bibitem{Emre2013} B. Emre, S. Y\"uce, E. Stern-Taulats, A. Planes, S. Fabbrici, F. Albertini, and L. Ma\~nosa, J. Appl. Phys. {\bf 113}, 213905.


\bibitem{Richardson1973} M. J. Richardson, D. Melville, and J. A. Ricodeau, Phys. Lett.  {\bf 46A}, 153 (1973).

\bibitem{Yuce2012} S. Y\"uce, M. Barrio, B. Emre, E. STern-Taulats, A. Planes, J. L. Tamarit, Y. Mudryk, K. A. Gschnmeidner, V. K. Pecharsky, and L. Ma\~nosa, Appl. Phys. Lett. {\bf 101}, 071906 (2012).

\bibitem{Casanova2005} F. Casanova, A. Labarta, X. Batlle, F. J. P\'erez-Reche, E. Vives, L. Ma\~nosa, and A. Planes, Appl. Phys. Lett. {\bf 86}, 262504 (2005).

\bibitem{SternTaulats2014} E. Stern-Taulats, P. O Castillo-Villa, L. Ma\~nosa, C. Frontera, S. Pramanick, S. Majumdar, and A. Planes, J. Appl. Phys. {\bf 115}, 173907 (2014).

\bibitem{Casanova2004} F. Casanova, A. Labarta, X. Batlle, J. Marcos, L. Ma\~nosa, A. Planes, and S. de Brion, Phys. Rev. B {\bf 69}, 104416 (2004).

\bibitem{Recarte2010} V. Recarte, J.I. Pe\'erez-Landaz\'abal, S. Kustov, and E. Cesari, J. Appl. Phys. {\bf 107}, 053501 (2010).

\bibitem{Casanova2002} F. Casanova, X. Batlle, A. Labarta, J. Marcos, L. Ma\~nosa, and  A. Planes, Phys. Rev. B {\bf 66}, 100401 (2002).



\bibitem{Manosa2014} L. Ma\~nosa, E. Stern-Taulats, A. Planes, P. Lloveras, M. Barrio, J. L. Tamarit, B. Emre, S. Y\"uce, S. Fabbrici, and F. Albertini,
phys. stat sol. (b), doi 10.1002/pssb.201350371.

\bibitem{Aksoy2009} S. Aksoy, M. Acet, P.P. Den, L. Ma\~nosa, and A. Planes, Phys. Rev. B {\bf 79}, 212401 (2009).

\bibitem{Recarte2012} V. Recarte, J.I. P\'erez-Landaz\'abal, V. S\'anchez-Alarcos, V. Zablotskii, E. Cesari, and S. Kustov, Acta mater. {\bf 60}, 3168 (2012).

\bibitem{Planes2001} A. Planes, and L. Ma\~nosa, Sol. Stat. Phys. {\bf 55}, 159 (2001).

\bibitem{Siewert2012} M. Siewert, M. E. Gruner, A. Hucht, H.C. Herper, A. Dannenberg, A. Chakrabarti, N. Singh, A. Arr\'oyave, and P. Entel, Adv. Engn. Mater. {\bf 14}, 530 (2012).

\bibitem{Liu2012} J. Liu, T. Gotschall, K.P. Skokov, J.D. Moore, and O. Gutfleisch, Nature Mater. {\bf 11}, 620 (2012). 

\end{thebibliography}

\end{document}